\documentclass[aps,prl,twocolumn,superscriptaddress,showpacs,floatfix]{revtex4}

\usepackage{graphicx}                           
\usepackage{amsmath}
\usepackage{amssymb}
\usepackage{amsfonts}
\usepackage{amstext}

\usepackage{hyperref}
\usepackage[figure,table]{hypcap}               
\hypersetup{
pdftitle = {},
pdfsubject = {},
pdfauthor = {},
pdfkeywords = {},
pdfcreator = {},
pdfproducer = {LaTeX with hyperref},
colorlinks = true,
linkcolor = red,
anchorcolor = red,
citecolor = red, 
filecolor = red, 
menucolor = red, 
pagecolor = red, 
urlcolor  = red,
breaklinks = true,
pdfstartview = FitV,
pdfhighlight = /I,
pdfpagelayout = OneColumn,
hypertexnames=true
}

\begin{document}

\title{Nonequilibrium Spin Dynamics in the Ferromagnetic Kondo Model}

\author{Andreas Hackl}
\affiliation{Institut f\"{u}r Theoretische Physik, Universit\"{a}t zu
K\"{o}ln, Z\"{u}lpicher Str.~77, 50937 K\"{o}ln, Germany}
\affiliation{Department of Physics, Harvard University, Cambridge, Massachusetts 02138, USA}
\author{David Roosen}
\affiliation{Institut f\"{u}r Theoretische Physik, Goethe Universit\"{a}t, 60438 Frankfurt/Main, Germany}
\author{Stefan Kehrein}
\affiliation{Arnold Sommerfeld Center for Theoretical Physics and CeNS, Department Physik, Ludwig-Maximilians-Universit\"{a}t, 80333 M\"{u}nchen, Germany}
\author{Walter Hofstetter}
\affiliation{Institut f\"{u}r Theoretische Physik, Goethe Universit\"{a}t, 60438 Frankfurt/Main, Germany}

\date{March 5, 2009}

\begin{abstract}
    Motivated by recent experiments on molecular quantum dots we
    investigate the relaxation of pure spin states when coupled to metallic
    leads. Under suitable conditions these systems are well described by a
    ferromagnetic Kondo model. Using two recently developed theoretical
    approaches, the time-dependent numerical renormalization group and an
    extended flow equation method, we calculate the real-time evolution of
    a Kondo spin into its partially screened steady state. We obtain exact
    analytical results which agree well with numerical implementations of
    both methods. Analytical expressions for the steady state magnetization
    and the dependence of the long-time relaxation on microscopic
    parameters are established. We find the long-time relaxation process
    to be much faster in the regime of anisotropic Kondo couplings. The
    steady state magnetization is found to deviate significantly from its
    thermal equilibrium value. 
\end{abstract}


\maketitle

\begin{figure}
    \includegraphics[width = 0.8 \columnwidth]{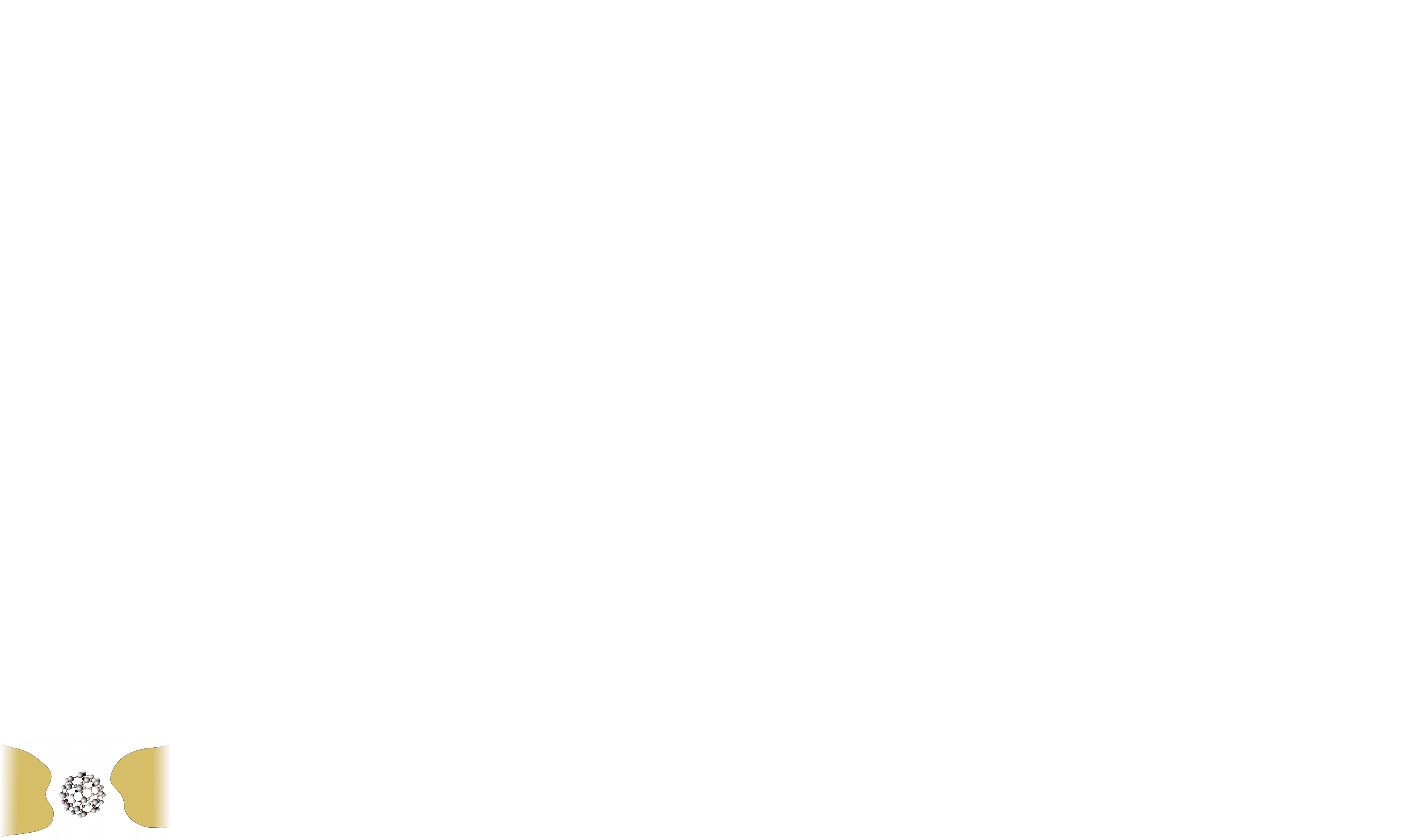}
    \caption{\label{fig:scheme} Sketch of a C$_{60}$-molecule coupled to
    metallic leads.
    }
\end{figure}

{\em Introduction.}
Recently it has become experimentally feasible to trap isolated single
molecules in nanogaps forming transistor geometries. In such molecular
quantum dots a variety of interesting new phenomena have been observed. In
case of a single C$_{60}$-molecule attached to metallic leads (sketched in
Fig.~\ref{fig:scheme}) the quantum phase transition between a singlet and a
triplet eigenstate of the molecule has been studied in detail
\cite{Roch2008,Hofstetter2001,Pustilnik2001}. In particular, if the
isolated molecule is prepared in the triplet configuration, its spin is
partially screened by the conduction band. In this case the resulting
effective exchange interaction between the residual spin and the conduction
band is known to be ferromagnetic \cite{Nozieres1980, Roosen2008}.

Replacing C$_{60}$ by a single-molecule magnet (SMM) such as Mn$_{12}$
gives rise to even more complex quantum impurity physics
\cite{Heersche2006}. As a result of magnetic anisotropy induced by
spin-orbit coupling, the large intrinsic spin of the SMM tends to align
along the easy axis of the molecule. This gives rise to an energy barrier
which suppresses magnetization reversal and makes SMMs promising candidates
for applications such as high-density magnetic storage and quantum
information processing \cite{Leuenberger2001}. When coupled to metallic
leads, the SMM can be described by an effective Kondo Hamiltonian with
anisotropic exchange coupling between the impurity spin and the conduction
band \cite{Romeike2006,Romeike2006a}
\begin{align}
    \mathcal{H} & = \sum_{\vec{k} \sigma} \varepsilon_{\vec{k}}
    c_{\vec{k} \sigma}^{\dagger} c_{\vec{k} \sigma}^{\phantom{\dagger}}
    + \frac{J_{\perp}}{2} \sum_{\vec{k} \vec{k^{\prime}}} ( c_{\vec{k}
    \uparrow}^{\dagger} c_{\vec{k^{\prime}} \downarrow}^{\phantom{\dagger}}
    S^{-} + c_{\vec{k} \downarrow}^{\dagger} c_{\vec{k^{\prime}}
    \uparrow}^{\phantom{\dagger}} S^{+} ) \nonumber \\
    & + \frac{J_{\parallel}}{2} \sum_{\vec{k} \vec{k^{\prime}}} ( c_{\vec{k}
    \uparrow}^{\dagger} c_{\vec{k^{\prime}} \uparrow}^{\phantom{\dagger}} -
    c_{\vec{k} \downarrow}^{\dagger} c_{\vec{k^{\prime}}
    \downarrow}^{\phantom{\dagger}}) S_{z} + 
    g \mu_B h S_{z}.
    \label{eq:AKM_def}
\end{align}
In the cotunneling regime it has been shown \cite{Gonzalez2008} that the
exchange interaction is ferromagnetic, i.~e.~\mbox{$J_{\parallel} \ll
J_{\perp} < 0$}, if adding or subtracting an electron to the molecule
increases the spin of the SMM. Preparing the system in a well-defined spin
state and measuring the real-time spin dynamics can be achieved using
electrical \cite{Nowack2007} or optical \cite{Braun2005,Atature2007} field
pulses, albeit experimental challenges in applying these techniques to
molecular quantum dots still remain.

Although the Kondo Hamiltonian has been studied in great detail, the
ferromagnetic regime has often been neglected (an exception is the
investigation of the spatial equal-time spin-correlations of an
underscreened spin-1 impurity \cite{Borda2008}). In this letter we will
focus on two important questions arising in this context: By studying the
magnetization dynamics we will investigate how fast an initially polarized
spin will reduce its magnetization due to spin flip scattering. For the
antiferromagnetic Kondo model this question has been answered in
\cite{Lobaskin2008,Anders20052006}. Our analysis yields important
information about the dominant relaxation mechanism in related experiments.

One further important question regarding the relaxation process is the
nature of the final state of the quantum system. Since a pure state
remains pure under unitary time evolution, the complete system is not
expected to behave like an equilibrium state even at long times. In
equilibrium, the conduction electrons will not fully screen the spin, as is
well known for the ferromagnetic Kondo model \cite{Abrikosov1970}.
Instead, the coupling $J_\perp$ provides weak spin flip scattering which
renormalizes the magnetization of the impurity spin to some finite value in
presence of a symmetry-breaking infinitesimal magnetic field. In the case
of isotropic couplings and a spin $S$, this value is known as
\cite{Abrikosov1970}
\begin{equation}
    \langle S_z \rangle = S(1+ \frac{J\rho}{2} + O(J^2)),
    \label{magnetization}
\end{equation}
where $\rho$ is the density of states in the conduction band with support
$[-D,D]$ (we assumed $\rho=(2D)^{-1}$ and employ units in which 
$\hbar=k_B=D=1$ in the following). However, the fact that the system's final
state differs from thermal equilibrium does not necessarily imply that
local observables retain a memory of the initial preparation, and in fact
even for certain integrable systems the reduced density matrix of a local
subsystem is known to thermalize \cite{Cramer2008}. For the ferromagnetic
Kondo model, the low-energy spin-flip scattering rate renormalizes to zero,
leaving open the question whether in the steady state at long times the
impurity has finite magnetization or not. We will give a definite answer
to this question in this letter and show that the asymptotic nonequilibrium
magnetization at long times differs from the equilibrium value.
Therefore in our model information about the initial preparation of the
system is never completely lost, even for local observables.

{\em Methods.}
In recent years, several numerical approaches have been developed to
calculate real-time dynamics of quantum impurity systems
\cite{Werner2009,Anders20052006,Schollwock2006,Schoeller2000}. However, the
accuracy of numerical data is usually not sufficient to give precise
answers about the nature of the long-time decay, i.~e.~to identify
analytical laws for the long-time tails and steady state values. We
therefore use an analytical approach to identify the long-time behavior
and compare it against numerical calculations to validate our analytical
approximations. We first describe our analytical approach, before briefly
sketching the numerical technique.

Within a poor man's scaling analysis a ferromagnetic exchange coupling of
an impurity spin to a fermionic bath renormalizes to zero at the Fermi
energy. This allows perturbative renormalization techniques to accurately
describe the low-energy physics of such a system. In this context, a
powerful technique is the flow equation method as invented by Wegner
\cite{Wegner1994} and independently by G{\l}azek and Wilson
\cite{Glazek19931994}. In a recent modification of the original
flow equation method it has been shown that the underlying renormalization
scheme can be extended to calculate the real-time evolution of interacting
many-body systems \cite{Hackl2008,Hackl2009,Moeckel2008}. As a notable
feature, this approach allows to derive exact analytical results. For
further details of the flow equation approach and its application to the
Kondo problem we refer to Ref.~\cite{Kehrein2006}.

We briefly outline the main steps of the flow equation calculation. Details
of this calculation will be published elsewhere \cite{Hackl2009a}.
As usual, the impurity spin operator is first transformed by a sequence of 
infinitesimal unitary transformations. The flowing spin operator has the form 
\begin{equation}
    S_z(B)=h(B)S_z +\sum_{kk^\prime} \gamma_{k^\prime k}(B) :(S^+
    s_{k^\prime k}^- +S^- s_{k^\prime k}^+): ,
\end{equation}
where the initial form of the operator is obtained for $B=0$
\cite{Kehrein2006}. Here, the operators $s_{k^\prime k}^\pm$ are matrix
elements of the conduction electron spin density raising and lowering
operators. The coupling constants $h(B)$ and $\gamma_{k^\prime k}(B)$ obey
the flow equations
\begin{eqnarray}
    \frac{dh}{dB} &=& \sum_{kk^\prime} (\varepsilon_{k^\prime} -\varepsilon_k)
    J_{k^\prime k}^\perp (B) \gamma_{kk^\prime} (B)
    n(k^\prime)(1-n(k))\nonumber\\
    \frac{d\gamma_{k^\prime k}}{dB} &=& h(B)(\varepsilon_{k^\prime} -
    \varepsilon_k) J_{k^\prime k}^\perp (B) +O(J^2),
    \label{feqsz}
\end{eqnarray}
where $n(k)$ denotes the Fermi distribution function. In addition, the
flowing couplings $J_{k^\prime k}^\perp (B)$ and $J_{k^\prime
k}^\parallel(B)$ of the Hamiltonian enter, which have to be calculated
separately \cite{Kehrein2006}. The fixed point of the transformation is
reached in the limit $B \rightarrow \infty$, where we denote coupling
constants in this basis by a tilde, e.~g.~$\tilde{h}$. The Heisenberg
equation of motion of the impurity spin can be solved efficiently by
solving it first for the transformed impurity spin and reverting the
unitary flow afterwards \cite{Hackl2008,Hackl2009}. By solving the
equations of motion in the diagonal basis of the Hamiltonian, one avoids
secular terms that grow in an uncontrolled way with time and can obtain
controlled analytical results even for the asymptotic long-time behavior. 

In order to verify this semi-analytical approach we employ the recently
introduced time-dependent numerical renormalization group method (TD-NRG)
by Anders and Schiller \cite{Anders20052006}. Describing this method in detail
is beyond the scope of this letter. Let us only mention here that it is
tailored to calculate the response of an (arbitrary) quantum impurity
system to a sudden quench at time $t=0$ and is able to access the long
time-scales characteristic for Kondo physics. We refer the interested
reader to Ref.~\cite{Anders20052006} for more details on this method.

{\em Results.}
For our analytical calculations, we assume that the impurity spin is
prepared in the up state of the spin projection operator $S_z$ before the
thermalized conduction electron bath ($| FS \rangle$) is coupled 
to it, leading to a product initial state
\begin{equation}
    | \psi \rangle = | \uparrow \rangle \otimes | FS \rangle .
\end{equation}

At time $t=0$, the spin is coupled to the conduction electrons, which
would e.~g.~be realized by attaching metallic leads to the single molecule
magnet. In the following, we restrict our calculations to spin $S=1/2$ and
zero magnetic field. Numerical calculations which we performed for the dynamics of
larger spins $S$ all satisfied the trivial relation $\langle S_z(t)
\rangle= S \langle \sigma_{z}(t)\rangle$, which is exact up to $O(J^2
S^2)$ from higher order flow equations \cite{Hackl2009a}.
Here $S$ denotes the size of the spin and $\sigma_{z}$ is the spin operator for 
spin $S=1/2$. After solving the Heisenberg equation of motion for the operator 
$S_z$, the formal result for the magnetization reads
\begin{equation}
    \langle S_z(t)\rangle=\frac{\tilde{h}}{2} + \sum_{kk^\prime}
    \frac{\tilde{\gamma}_{kk^\prime}^2}{2} \bigl( e^{it(\varepsilon_k-\varepsilon_{k^\prime})}-\frac{1}{2} \bigr)
    n(k^\prime)(1-n(k)) . 
    \label{formalht}
\end{equation}

In the following, we will only discuss the purely quantum mechanical case,
corresponding to $T=0$. It turns out that the magnetization dynamics can be
fully understood by the energy dependency of the couplings
$\tilde{\gamma}_{kk^\prime}$, which are obtained from a solution to
Eq.~\eqref{feqsz}.

{\em Isotropic Kondo Model.}
Let us first investigate an isotropically coupled spin. In equilibrium,
perturbative scaling shows \cite{Hewson1997} that the isotropic coupling
$J$ logarithmically decreases upon reducing the half band width from $D$ to
some $\Lambda<D$, 
\begin{equation}
    J(\Lambda )= \frac{J}{1+ \rho J \ln \bigl(\frac{\Lambda}{D}\bigr)}.
    \label{isoflow}
\end{equation}
At the low energy fixed point, an infinitesimal magnetic field is
sufficient to polarize the free spin, leading to a finite magnetization
according to Eq.~\eqref{magnetization}.

In nonequilibrium, our results show that the magnetization saturates as
well, but to a different value than in equilibrium. Using the exact low
energy behavior of the couplings $\tilde{\gamma}_{kk^\prime}$ in
Eq.~\eqref{formalht}, the asymptotic behavior of the magnetization is
obtained as
\begin{equation}
    \langle S_z(t)\rangle = \frac{1}{2} \biggl(\frac{1}{\ln(t)
    -\frac{1}{\rho J}} + 1 + \rho J + O(J^2)\biggr).
    \label{isotail}
\end{equation}
This behavior can be understood from the logarithmic renormalization of the
coupling $J$, which directly enters the low energy flow of the couplings
$\gamma_{kk^\prime}$ via Eq.~\eqref{feqsz}.

The steady state magnetization $\langle S_z(t \rightarrow \infty)\rangle =
\frac{1}{2}(1+\rho J+O(J^2))$ therefore differs from the equilibrium value
as given by Eq.~\eqref{magnetization}. The reduction from full
polarization is $\rho J/2$, which is twice the equilibrium value. This can
be attributed to the fact that the nonequilibrium dynamics starts with an
impurity spin that is not dressed with a conduction band electron cloud: it
therefore relaxes to a smaller value of the magnetization as compared to
the dressed impurity spin in equilibrium.

\begin{figure}
    \includegraphics[width = 1.0 \columnwidth]{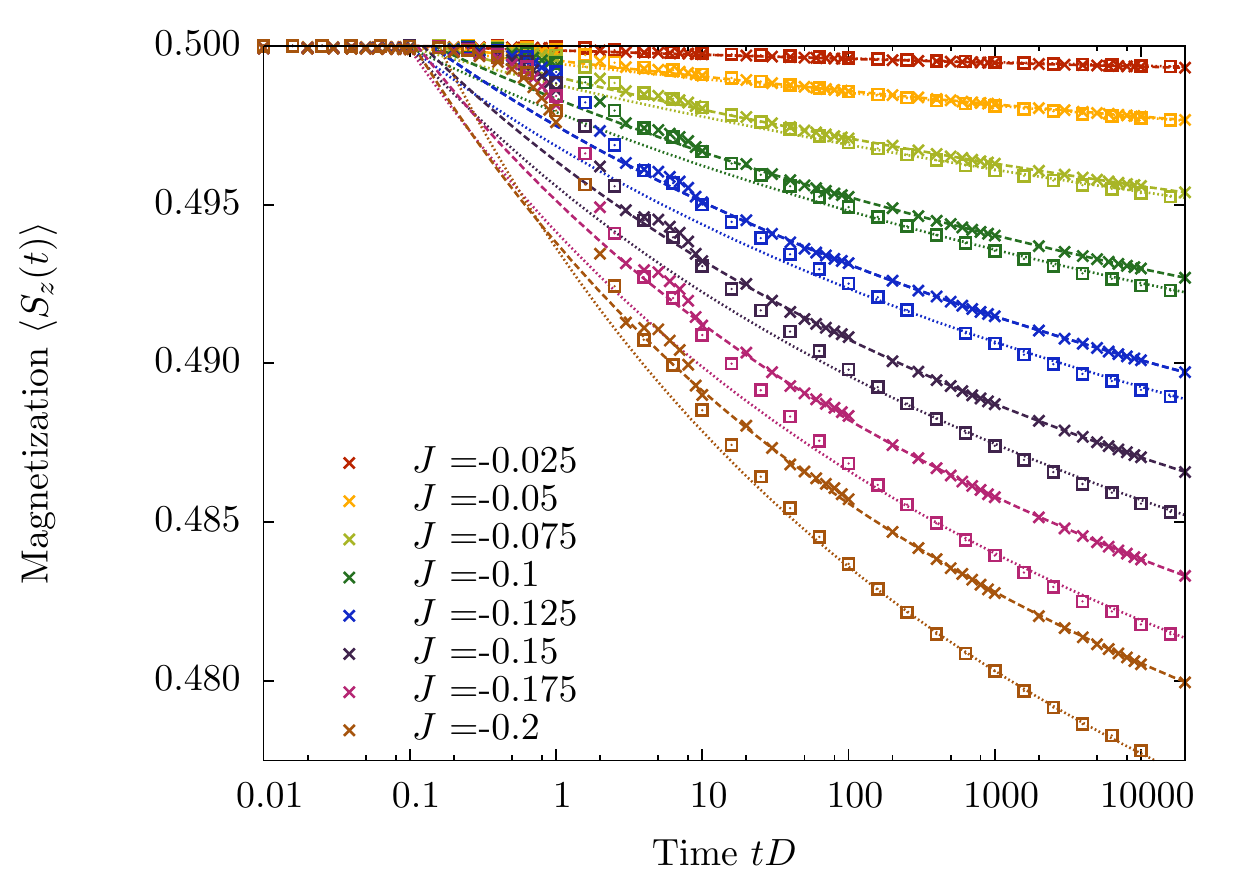}
    \caption{\label{fig:iso} 
    Results for the \emph{isotropic} ferromagnetic Kondo model ($\times$:
    TD-NRG data, $\Box$: flow equation data). Using our analytical
    result we fitted our data against $\langle S_{z} (t) \rangle = (1 +  a
    J_{\parallel} + ( \ln(t) - 1/(c J_{\parallel} ) )^{-1} ) /2$ using
    $a,c$ as fit parameters (lines).
    }
\end{figure}

A direct numerical solution of the flow equations allows to accurately
determine the relaxation process also at intermediate and short time
scales. Together with the analytical result from Eq.~\eqref{isotail}, this
calculation can be compared to TD-NRG calculations. Both methods yield
very good agreement up to time scales of order $t \approx 10^{4}$ where the
asymptotic logarithmic relaxation is clearly visible, see
Fig.~\ref{fig:iso}. Increasing deviation of the curves for larger coupling
strength $J$ can be explained by the $O(J^2)$ corrections to the flow
equation result, which we neglected. We checked that the relative deviation
of the two methods in terms of the quantity $\langle S_z (t)\rangle-1/2$ at
some large but fixed time indeed grows approximately linearly in $J$. A
fit of the TD-NRG curves and the numerical implementation of the flow
equation approach are in good agreement with the analytical result of
Eq.~\eqref{isotail}.

\begin{figure}
    \includegraphics[width = 1.0 \columnwidth]{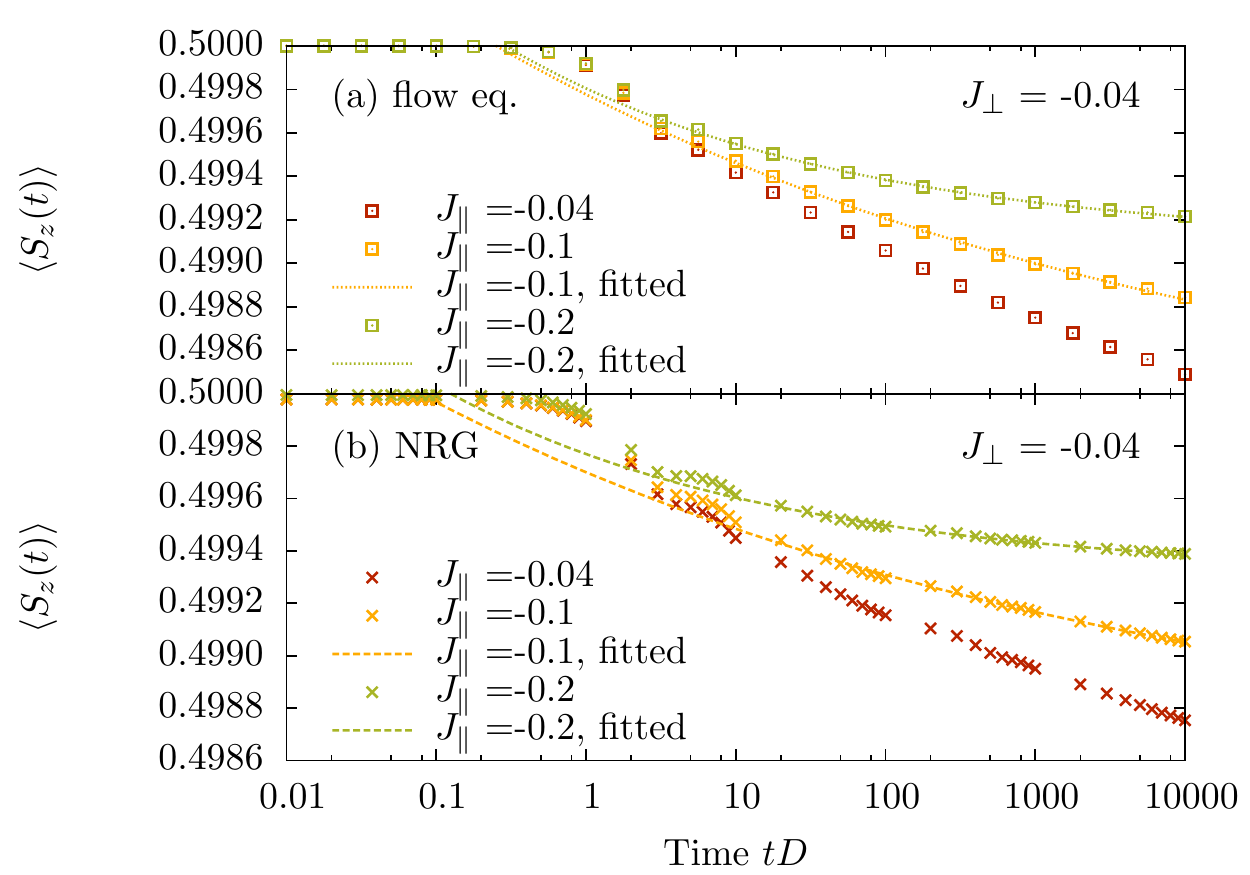}
    \caption{\label{fig:aniso} 
    For the \emph{anisotropic} ferromagnetic Kondo model our numerical
    findings coincide with our analytical results using both methods.
    Fitting our data against $\langle S_{z}(t) \rangle = a \cdot
    t^{-\sqrt{J_{\parallel}^{2} - J_{\perp}^{2}}} + c $ we found good
    agreement for the fit parameters $a,c$.
    }
\end{figure}

{\em Anisotropic Kondo Model.}
Studying the anisotropic Kondo model we restrict ourselves to the
experimentally relevant case $J_\perp > J_\parallel$ in the following.
From a poor man's scaling analysis \cite{Hewson1997}, it is known that the
coupling $J_\perp$ renormalizes as $J_\perp(\Lambda) \propto \Lambda^{\rho
\sqrt{J_\parallel^2-J_\perp^2 }}$ at low energies $\Lambda$. As in the
isotropic case, the flow equation analysis shows that this behavior
determines the asymptotic long-time relaxation of the spin, given
explicitly by the power law
\begin{equation}
    \langle S_z(t)\rangle= 0.5\biggl(1 -
    \frac{\alpha^2}{2\tilde{g}_\parallel} t^{2\tilde{g}_\parallel}
    +\frac{\alpha^2}{2\tilde{g}_\parallel}+O(J^2)\biggr),
    \label{anisotail}
\end{equation}
where $\tilde{g}_\parallel = -\rho \sqrt{J_{\parallel}^2-J_{\perp}^2}$. The
constant $\alpha$ derives from the scaling equations for $J_\perp$ and
$J_\parallel$. Numerical checks show that it can be replaced by $\alpha
\approx \rho J_\perp $ as long as $J_\parallel \lesssim 2 J_\perp$. In
comparison to the isotropic case, the power law decay of spin flip
scattering at low energies leads to much faster relaxation of the
magnetization, whereas the steady state magnetization is enhanced. This
behavior is reproduced by our numerical calculations shown in
Fig.~\ref{fig:aniso}. Again, our calculations showed that the steady state
magnetization $\langle S_z (t\rightarrow \infty)\rangle =1/2 + (\alpha^2/(4
\tilde{g}_\parallel)$ is reduced twice as much from full polarization than
in equilibrium. The analytical results are confirmed by numerical fits of
our data, see Fig.~\ref{fig:aniso}.
Let us point out that for the anisotropic Kondo model, our methods are
starting from slightly different initial states. Using the flow equation
approach one is restricted to a situation where the spin is initially
completely decoupled from the fermionic bath. On the other hand, stability
of the TD-NRG algorithm in the anisotropic model requires preparing the
polarized spin at time $t<0$ by applying a large magnetic field, while
still allowing for a small exchange coupling to the metallic leads. The
same long-time power-law relaxation was obtained with both methods.
However, the slightly different initial states used in both methods become
significant on short and intermediate time scales.

{\em Conclusions.}
We employed two different methods to analyze the real-time evolution of a
ferromagnetically coupled Kondo spin, which is initially prepared in a
polarized state. Exact analytical results for the long-time behavior of
the magnetization were obtained for two different situations. For the
isotropic ferromagnetic Kondo model, the long-time relaxation is
logarithmic in time, whereas anisotropic couplings lead to a power-law
decay at large times. 
Furthermore, exact analytical results for the asymptotic nonequilibrium
magnetization were presented, which differ from the equilibrium
magnetizations. They confirm that the local quantum impurity retains a
memory of the initial preparation for asymptotically large times. This is
due to the combined effect of nonequilibrium preparation and ergodicity
breaking already in the equilibrium system.

\begin{acknowledgments}
We have benefited from discussions with M.~Vojta, A.~Rosch and M.~Garst.
A.~H.~acknowledges support through SFB 608, SFB/TR12 and FG 960.
S.~K.~ acknowledges support through SFB/TR 12 and FG 960 of the Deutsche
Forschungsgemeinschaft (DFG), the Center for Nanoscience (CeNS) Munich, and
the German Excellence Initiative via the Nanosystems Initiative Munich
(NIM).
\end{acknowledgments}

\end{document}